# Ambient-pressure superconductivity onset above 40 K in bilayer nickelate ultrathin films

Guangdi Zhou[1†], Wei Lv[1†], Heng Wang[1†], Zihao Nie[1†], Yaqi Chen[1], Yueying Li[1], Haoliang Huang[1,2], Weiqiang Chen[1,2], Yujie Sun[1,2], Qi-Kun Xue[1,2,3*], Zhuoyu Chen[1,2*]

[1]Department of Physics and Guangdong Basic Research Center of Excellence for Quantum Science, Southern University of Science and Technology, Shenzhen 518055, China
[2]Quantum Science Center of Guangdong-Hong Kong-Macao Greater Bay Area, Shenzhen 518045, China
[3]Department of Physics, Tsinghua University, Beijing 100084, China
†These authors contributed equally.
*E-mail: xueqk@sustech.edu.cn, chenzhuoyu@sustech.edu.cn

The discovery of bilayer nickelate superconductors under high pressure has opened a new chapter in high-transition temperature (high-$T_C$) superconductivity[1-8]. Here, we report ambient-pressure superconductivity onset above the McMillan limit (40 K) in bilayer nickelate epitaxial ultrathin films. Three-unit-cell (3UC) thick La$_{2.85}$Pr$_{0.15}$Ni$_2$O$_7$ single-phase-crystalline films are grown using the gigantic-oxidative atomic-layer-by-layer epitaxy (GOALL-Epitaxy) on SrLaAlO$_4$ substrates[9]. Resistivity measurements and magnetic-field responses indicate onset $T_C$ = 45 K. The transition to zero resistance exhibits characteristics consistent with a Berezinskii–Kosterlitz–Thouless (BKT)-like behavior, with $T_{BKT}$ = 9 K. Meissner diamagnetic effect is observed at $T_M$ = 8.5 K via a mutual inductance setup, in agreement with the BKT-like transition. In-plane and out-of-plane critical magnetic fields exhibit anisotropy. Scanning transmission electron microscopy (STEM) images and X-ray reciprocal space mappings (RSMs) show that the films maintain a tetragonal phase with coherent epitaxial compressive strain ~2% in the NiO$_2$ planes relative to the bulk. Our findings pave the way for comprehensive investigations of nickelate superconductors under ambient pressure conditions and for exploring superconductivity at higher transition temperature through strain engineering in heterostructures.

# Main

High-$T_C$ superconductivity was first discovered in cuprates that consist of stacked superconducting $CuO_2$ planes[10-13]. In each Cu ion within the Cu-O planes, approximately 9 electrons occupy the $3d$ orbitals, with the $3d_{x2-y2}$ orbital dominating near the Fermi level. The infinite-layer nickelates with superconducting $NiO_2$ planes[14], which are structurally analogous to cuprates and share the $3d^9$ configuration, exhibit a comparable electronic structure near the Fermi level, though with weaker oxygen hybridization, placing them closer to the Mott-Hubbard regime[15-24]. By contrast, the recently discovered bilayer nickelate superconductors possess a nominal $3d^{7.5}$ configuration, resulting in a distinct electronic structure compared to both cuprates and infinite-layer nickelates[1,7]. Structurally, unlike cuprates and infinite-layer nickelates that lack apical oxygen atoms between the active superconducting planes, the layered nickelates feature apical oxygens between the adjacent $NiO_2$ layers, suggesting a possible role of the $3d_{z2}$ orbital[1,6,25-30]. Despite all these differences, bilayer nickelates can host $T_C$ as high as the liquid nitrogen temperature, although only under high-pressure conditions, marking them as a crucial system for understanding the mechanism of high-$T_C$ superconductivity.

Superconductivity in bilayer nickelate $La_3Ni_2O_7$ was first observed at pressures exceeding 14 GPa, with resistance approaching zero[1,2]. Subsequent studies achieved zero-resistance states in both single-crystalline and polycrystalline $La_3Ni_2O_7$ samples under hydrostatic pressure, though the superconducting volume fraction remained low[3,4]. The presence of structural polymorphs, including various Ruddlesden-Popper variants, poses significant challenges, and obscures the identification of the phase truly responsible for superconductivity[31-34]. More recent efforts, such as substituting one-third of La with Pr to form $La_2PrNi_2O_7$, have successfully reduced phase intergrowth, improved the bilayer structure purity, which enhance greatly the superconducting volume fraction[5]. Nevertheless, the reliance on high-pressure conditions and the persistence of residual phase inhomogeneities continue to limit comprehensive experimental investigations into underlying superconducting mechanism and potential technological applications of bilayer nickelates. Crucially, whether high $T_C$ above the 40-K McMillan limit is possible under ambient pressure remains a significant open question. Here, through substrate-induced compressive strain on epitaxial ultrathin films with rare-earth element substitutions, we have achieved ambient-pressure superconductivity with transition onset at 45 K in a pure-phase-crystalline bilayer nickelate, thus opening a direct pathway for experimental studies of its superconducting properties.

$La_{2.85}Pr_{0.15}Ni_2O_7$ bilayer nickelate ultrathin films are grown on treated (001)-oriented $SrLaAlO_4$ substrates (Fig. 1a) by GOALL-Epitaxy[9]. In contrast to oxide molecular beam epitaxy (OMBE) or pulsed laser deposition (PLD), GOALL-Epitaxy uses different laser-ablation targets for different atomic layer under strong ozone oxidation environment (Extended Data Fig. 1). For the growth of a $La_{2.85}Pr_{0.15}Ni_2O_7$ thin film, $La_{0.95}Pr_{0.05}O_x$ and $NiO_x$ targets are alternatively ablated according to a set sequence to form periodic stackings of $La_{0.95}Pr_{0.05}O$-$NiO_2$-$La_{0.95}Pr_{0.05}O$-$NiO_2$-$La_{0.95}Pr_{0.05}O$. Each period corresponds to half unit cell of the lattice structure. The growth process is monitored *in situ* with reflective high energy electron diffraction (RHEED) under an oxidative condition mixing 3-5 Pa of purified ozone delivered through a 9-mm-diameter nozzle positioned near the sample (~3.5 cm distance) and 7 Pa of background oxygen (Extended Data Fig. 2). During growth the epitaxial strain is coherently maintained within the 3UC thickness (Extended Data Fig. 3). The stoichiometry precision is better than 1%. Samples are cooled down at 100°C/min under the same ozone flow, and extracted for X-ray diffraction (XRD) measurements and Pt electrode depositions with magnetron sputtering. To achieve optimal superconductivity, post-annealing in the growth chamber for 30 min at 575 °C, with purified ozone of 15 Pa (flow rate ~10 sccm) continuously injected through the same nozzle, is critical. The samples are then cooled in the same ozone environment at 100°C/min until reaching near room temperature. After shutting off the ozone flow, the samples are promptly transferred from the growth chamber to the load-lock chamber. Once the load-lock gate valve is closed, pure oxygen is immediately introduced to vent the load-lock chamber, ensuring a consistently strong oxidation environment to stabilize the superconducting samples. Extended Data Fig. 4 shows resistivity measurement before and after post-annealing. Post-annealing does not alter structural characteristics (Extended Data Fig. 5).

An 3-unit-cell (3UC) $La_{2.85}Pr_{0.15}Ni_2O_7$ film on $SrLaAlO_4$ substrate grown and annealed with optimal conditions exhibits superconductivity with onset $T_C$ of 45 K

(Fig. 1b). Films grown and annealed under sub-optimal conditions (e.g. different temperatures) are shown in Extended Data Fig. 6. The resistivity-temperature (*R-T*) curve undergoes a two-step resistive drop, a behavior observed previously in disordered two-dimensional (2D) superconductors that can be described as Josephson-coupled superconducting puddles[35-37]. Only temperatures below 200 K are shown, as higher temperatures under rough vacuum conditions were found to cause oxygen loss in the samples, leading to the deterioration of superconducting properties. The onset $T_C$ is defined as the temperature at which resistivity begins to deviate from a linear fit between 50 K and 60 K. The determined onset $T_C$ = 45 K also coincides with the temperature where resistivity exhibits obvious magnetic field responses, such as at 14 T (Fig. 1c). Above onset $T_C$, the *R-T* curve exhibits temperature dependence deviates from a linear behavior previously observed in pressured bulks[3] (Fig. 1d). This may indicate different disorder or doping levels in these different systems. Phonon scatterings is also an important contributor in this temperature regime. Below 20 K, resistivity reaches zero gradually, close to a Berezinskii–Kosterlitz–Thouless (BKT)-like behavior (Fig. 1e), as previously seen in 2D/interfacial superconductors[38,39]. A linear fit to the $(d\ln(R)/dT)^{-2/3}$-versus-*T* curve yields $T_{BKT}$ = 9 K (Fig. 1f). The current-voltage (*I-V*) characteristics exhibit power-law behavior similar to a BKT transition[40] (Extended Data Fig. 7). By contrast, a 3UC La$_3$Ni$_2$O$_7$ film grown on the same substrate shows a small resistive drop at ~10 K, without reaching zero resistance (Extended Data Fig. 8). Given the ultrathin thickness of the film, interfacial disorder may play a key role in affecting superconducting phase coherence across the whole film. This may be related to that $T_{BKT}$, corresponding to phase coherence, is significantly lower than onset $T_C$, which is linked more closely to Cooper pairing.

Samples with and without Pr substitutions display Hall effect behaviors (Fig. 1h, raw data) markedly different from typical cuprates[41,42] (like La$_{1.85}$Sr$_{0.15}$CuO$_4$) and infinite-layer nickelates[14-16] (like Nd$_{0.8}$Sr$_{0.2}$NiO$_2$/SrTiO$_3$). In particular, absolute values of the Hall coefficients of both La$_3$Ni$_2$O$_7$ and La$_{2.85}$Pr$_{0.15}$Ni$_2$O$_7$ samples are around 2 orders of magnitudes lower. A single band model yields approximately 10 electrons or holes per Ni at 40 K, for La$_3$Ni$_2$O$_7$ or La$_{2.85}$Pr$_{0.15}$Ni$_2$O$_7$, respectively, which is unphysically large. In the multi-channel case, without access to ultra-high magnetic fields, the Hall effect remains linear (Extended Data Fig. 9), and the near-zero Hall coefficients indicate the coexistence of electron- and hole-like carriers.

However, the role of channel-dependent mobility in the Hall effect precludes quantitative determination of electron or hole carrier concentrations, though their coexistence is evident. These results suggest the multiband nature at the Fermi level.

The Meissner diamagnetic effect is demonstrated using a mutual inductance setup, with the driving and pickup coils aligned vertically above and below the thin film sample (Fig. 2a). A diamagnetic signal is observed below $T_M = 8.5$ K (Meissner temperature), which coincides with $T_{BKT}$ extracted from the $R$-$T$ curve within the error margin. This signifies the entrance of a true superconducting state, occurring at a temperature comparable to that of infinite-layer nickelates[14]. We extracted penetration depth $\lambda$ in the order of µm as a function of temperature, shown in Extended Data Fig. 10. To further elucidate the dimensionality of the superconductivity, in Fig. 2b and 2c we show the $R$-$T$ curves with varied out-of-plane and in-plane magnetic fields for the 3UC $La_{2.85}Pr_{0.15}Ni_2O_7$/$SrLaAlO_4$ sample (see Extended Data Fig. 11 for another onset-$T_C$-32-K sample). A strong anisotropy of magnetic response is seen. Out-of-plane and in-plane critical fields $B_c$'s are extracted, as defined by 90% and 50% normal state resistivity $R_N$, and plotted as functions of temperature (Fig. 2d). For the out-of-plane fields $B_c^\perp$'s, linear fits yield a zero-temperature critical fields of 68 T and 29 T, for 90% and 50% cases, respectively. In contrast, the in-plane fields $B_c^\parallel$'s are fitted using the 2D Ginzburg-Landau formula[43], resulting in a zero-temperature critical fields of 119 T and 71 T, for 90% and 50% cases, respectively. Note that the critical field anisotropy is less pronounced than what is typically observed in many 2D superconductors. However, the distinct dichotomy of linear and nonlinear temperature dependence in the out-of-plane and in-plane critical fields is hallmark of 2D superconductors[35,38,44-46], which is different from the behavior observed in infinite-layer nickelates[43,47-49]. Superconducting thickness can be estimated using the formula $d_{SC} = \sqrt{(6\phi_0 B_c^\perp/\pi B_c^{\parallel 2})}$, yielding $d_{SC} = 4 \pm 3$ nm. Considering the 90% critical fields as the pairing breaking upper critical fields $B_{c2}$, coherence lengths can be estimated using the relation $\xi_0 = \sqrt{(\phi_0/2\pi B_{c2})}$, yielding in-plane coherence length $\xi_0^\parallel = 2.2$ nm and out-of-plane coherence length $\xi_0^\perp = 1.7$ nm. These coherence lengths are comparable to the out-of-plane lattice constant and the extracted $d_{SC}$. Setting a 14 T magnetic field, $R$-$T$ curves are measured with varied rotation angle $\beta$. Temperatures at which resistivity reaches 50% of $R_N$ ($T_{50\%}$) as a function of $\beta$ is plotted in Fig. 2e. The $T_{50\%}$-$\beta$ curve can be well fitted with the Tinkham model for 2D superconductors[50], while the Ginzburg-Landau model for anisotropic 3D superconductor[51] fails to capture the cusp feature around zero angle. Since the extracted length scales are order-of-magnitude estimates, the magnetic field response data suggest that the bilayer nickelate ultrathin film superconductivity resides in a regime where $\xi_0 \sim d_{SC} \ll \lambda$.

Scanning transmission electron microscope (STEM) and atomically-resolved energy-dispersive X-ray spectroscopy (EDS) images reveal key structural characteristics of a superconducting 3UC $La_{2.85}Pr_{0.15}Ni_2O_7$/$SrLaAlO_4$ sample (Fig. 3). The spacing between two adjacent Ni-O planes is $4.09 \pm 0.02$ Å. At the interface, the substrate undergoes reconstructions most likely during the substrate treatments, forming an $AlO_2$ bilayer (specifically LaO-$AlO_2$-(La,Sr)O-$AlO_2$-(La,Sr)O), which acts as a template for epitaxial growth. The film growth begins with a LaO layer and follows a sequence of (La,Pr)O-$NiO_2$-(La,Pr)O-$NiO_2$-(La,Pr)O. The atomically sharp interface with no detectable interdiffusion between Ni and Al and the coherent epitaxial strain ensure that superconducting properties are preserved within ultrathin 3UC films near the interface. Interestingly, Sr diffusion from the substrate into the first unit cell of the film is observed. Since Sr substitution for La typically introduces hole doping, this may imply a thinner superconducting layer (assuming hole-doping is important) relative to the total film thickness, despite the single-phase structure observed throughout the film. The smaller ionic radius of $Pr^{3+}$ compared to $La^{3+}$ may have facilitated the diffusion of $Sr^{2+}$ (Extended Data Fig. 12), possibly enhancing superconductivity. The oxygen K edge of electron energy loss spectroscopy (EELS) shows $p$-$d$ hybridization in the film (Extended Data Fig. 13).

Figure 4 presents the X-ray diffraction (XRD) along the out-of-plane direction and reciprocal space mappings (RSMs) for $La_{2.85}Pr_{0.15}Ni_2O_7$ and $La_3Ni_2O_7$ films. The XRD shows a series of well-defined and diffraction peaks, indicating high crystalline quality even at ultrathin film regime. The out-of-plane lattice constant of the superconducting $La_{2.85}Pr_{0.15}Ni_2O_7$ film on $SrLaAlO_4$ substrate is calculated 20.74 Å, ~1% elongated compared to bulk (out-of-plane lattice constant is 20.519 Å). The non-superconducting $La_3Ni_2O_7$ film grown on $LaAlO_3$ substrate with a larger in-plane lattice constant, showing slightly smaller out-of-plane lattice constant, implying the possible role of sufficient straining for superconductivity. The presence of Kiessig fringes around the main diffraction peaks confirms that the sample has an atomically smooth surface and interface (also seen in atomic force microscope (AFM) images in Extended Data Fig. 14). X-ray reflectivity (XRR) measurements yield film thickness of ~6.6 nm for 3UC $La_{2.85}Pr_{0.15}Ni_2O_7$ film on $SrLaAlO_4$ substrate, in agreement with the expectation from growth within error margins (Extended Data Fig. 15). The RSMs of the superconducting 3UC $La_{2.85}Pr_{0.15}Ni_2O_7$ film along four different in-plane directions exhibit consistent $q_z$ values, providing evidence that the film retains a tetragonal phase. Alignment of $q_x$ between the film and the substrate demonstrates a uniform and coherent strain state across the film. In another word, the film has the same in-plane lattice constant as the $SrLaAlO_4$ substrate (3.75 Å), corresponding to a compressive strain of ~2% compared to bulk (in-plane lattice constant = 3.832 Å). Due to coherent straining for these ultrathin films, both 3UC $La_{2.85}Pr_{0.15}Ni_2O_7$ and 3UC $La_3Ni_2O_7$ films grown on $SrLaAlO_4$ substrates display nearly identical in-plane lattice constants within errorbars (3UC $La_3Ni_2O_7$/$SrLaAlO_4$ case shown in Extended Data Fig. 16).

In this work, we have achieved superconductivity in single-phase-crystalline bilayer nickelate ultrathin films at ambient pressure with transition onset above the McMillan limit. Using the GOALL-Epitaxy method, we grew high-quality, epitaxially strained 3-unit-cell $La_{2.85}Pr_{0.15}Ni_2O_7$ films on $SrLaAlO_4$ substrates with atomically smooth surfaces and interfaces. The ultrathin film shows an onset $T_C$ of 45 K and a BKT-like transition behavior with $T_{BKT}$ = 9 K, aligning with the Meissner temperature $T_M$. Magnetic responses reveal anisotropy in critical fields. Our work allows experimental investigations of the superconductivity mechanism with enhanced feasibility and lays a foundation for achieving transition temperature higher than 77 K (liquid nitrogen boiling temperature) in nickelates at ambient pressure.

## Methods

**Growth of $La_3Ni_2O_7$ and $La_{2.85}Pr_{0.15}Ni_2O_7$ films.** All samples were prepared using the gigantic-oxidative atomically layer-by-layer epitaxy (GOALL-Epitaxy) method. In the growth of $La_3Ni_2O_7$ and $La_{2.85}Pr_{0.15}Ni_2O_7$ films, which features a stacking block structure of (La,Pr)O-NiO$_2$-(La,Pr)O-NiO$_2$-(La,Pr)O, alternating laser ablation of LaO$_x$/La$_{0.95}$Pr$_{0.05}$O$_x$ and NiO$_x$ targets is employed to achieve atomic-layer-by-layer epitaxy. Calibrating the stoichiometry relies on determining both the precise number of laser pulses and energy required to ablate a target and deposit one monolayer on substrate. In a typical growth process, the number of pulses used to ablate LaO$_x$/La$_{0.95}$Pr$_{0.05}$O$_x$ or NiO$_x$ targets is between 100 and 150 pulses, providing a stoichiometric precision greater than 1%.

LaO$_x$, NiO$_x$, and La$_{0.95}$Pr$_{0.05}$O$_x$ targets were prepared by sintering LaO, NiO, or stoichiometric mixtures of LaO and PrO powders at 1100 °C for 6 hours. The growth temperature and laser fluence for all samples were set to be ~750 °C and 1.4 – 1.8 J/cm$^2$. The deposition atmosphere consists of two components: ozone and oxygen, with partial pressures of 3 Pa and 7 Pa, respectively. After the deposition, the samples were cooled down at 100°C/min in the same environment to improve experimental efficiency. 50°C/min and 60°C/min were also tested, with no significant differences in the crystalline quality observed.

**Post-annealing.** Post-annealing was performed for 30 min at 575°C in the growth chamber for optimal superconductivity, with purified ozone of ~15 Pa (flow rate ~10 sccm) delivered through a 9-mm-diameter nozzle. The samples were cooled in the same ozone environment at 100°C/min. The samples were extracted from the chamber until they reached near room temperature. Annealing has minimal effect on crystalline quality (Extended Data Fig. 17). Sample temperatures in the growth chamber were measured by pyrometer from the back of the inconel600 sample holder.

**Substrate preparation.** For clean surface preparation of the SrLaAlO$_4$ (001) substrates (MTI, China), annealing was performed by putting a LaAlO$_3$ substrate on a SrLaAlO$_4$ substrate face to face in the atmospheric or 1 atm oxygen condition at 1030 °C or 1080 °C for 2 hours. Treated substrates may exhibit either double or single termination surfaces, which do not significantly impact the film growth, strain, or lattice structure. Oxygen stoichiometry in the films is predominantly controlled during post-annealing.

**Low-temperature transport measurements.** Electric transport measurements were performed in a closed-cycle helium-free system (base temperature ~1.8K).

Hall bar electrodes with Pt were evaporated by magnetron sputtering with a prepatterned hard shadow mask covered on 5×5 mm samples. The four terminal electrical measurements were carried out through the standard lock-in technique with an AC current of 0.5 μA (13 Hz) for resistance measurements and 10 μA for Hall measurements. The transition temperatures under different magnetic fields are extracted from the resistivity versus magnetic field curves shown in Fig. 2a and 2b. The normal-state resistivity $R_N$ at a certain temperature is determined by linearly extrapolated from the 50–60 K range. Samples experience oxygen loss in vacuum above 200K and in 1 atom oxygen at ambient temperature, which can be characterized by resistance increase (Extended Data Fig. 18). Meanwhile, lattice structure characterized by XRD do not show noticeable difference (Extended Data Fig. 19).

**Mutual inductance measurements.** The superconducting thin film samples were placed tightly between the collinear driving and pickup coils. The driving coil and pick-up coil are both made using 30 μm diameter wires and have identical specifications. Each coil has an outer diameter of 1.5 mm, an inner diameter of 0.5 mm, a height of approximately 2.5 mm, and 800 turns, resulting in a self-inductance of 1 mH. The driving coil was driven with a 20 kHz 5μA alternating current. The voltage across the pickup coil is measured by a lock-in amplifier.

**Scanning transmission electron microcopy (STEM).** STEM annular bright field (ABF) and HAADF imaging of superconducting $La_{2.85}Pr_{0.15}Ni_2O_7$ and non-superconducting $La_3Ni_2O_7$ films were photographed using a FEI Titan Themis G2 at 300 kV, with a double spherical-aberration corrector (DCOR) and a high-brightness field-emission gun (X-FEG) with a monochromator installed onto this microscope. The inner and outer collection angles for the STEM images (β1 and β2) were 48 and 200 mrad, respectively, with a semi-convergence angle of 25 mrad. The beam current was about 80 pA for HAADF imaging and the EDS chemical analyses. The cross-section STEM specimens were prepared using a FEI Helios 600i dual-beam FIB/SEM machine. Before extraction and thinning, electron beam-deposited platinum and ion beam-deposited carbon were used to protect the sample surface from ion beam damage. And EDS data were obtained using the Super X FEI System in STEM mode. All operations were done at room temperature.

**X-ray diffraction (XRD).** Crystallographic characterization of thin-film specimens was performed using an automated multipurpose X-ray diffractometer (SmartLab, Rigaku Corporation), encompassing θ-2θ scans and reciprocal space mappings (RSMs). After the in-plane sample alignment, the RSMs around (1 0 11), (0 1 11), (-1 0 11) and (0 -1 11) SrLaAlO$_4$ Bragg reflections were measured by rotating the phi axis using the Hypix-3000 2D detector.

## Acknowledgements

We acknowledge the discussions with Guang-Ming Zhang, Dao-Xin Yao, and Hongtao Yuan. This work was supported by the National Key R&D Program of China (No. 2022YFA1403100), the Natural Science Foundation of China (Nos. 92265112, 12374455, and 52388201), and Guangdong Provincial Quantum Science Strategic Initiative (No. GDZX2401004, GDZX2201001 & SZZX2401001).

## Author contributions

Q.K.X. and Z.C. supervised the project. Z.C. initiated the study and coordinated the research efforts. G.Z., W.L. and Z.N. performed thin film growth with Y.C.'s assistance. H.W. performed low-temperature measurements and analysis. Y.L. and H.H. contributed on STEM analysis. H.H. contributed on XRD analysis. W.C. provided theoretical support. Z.C. wrote the manuscript with input from all other authors.

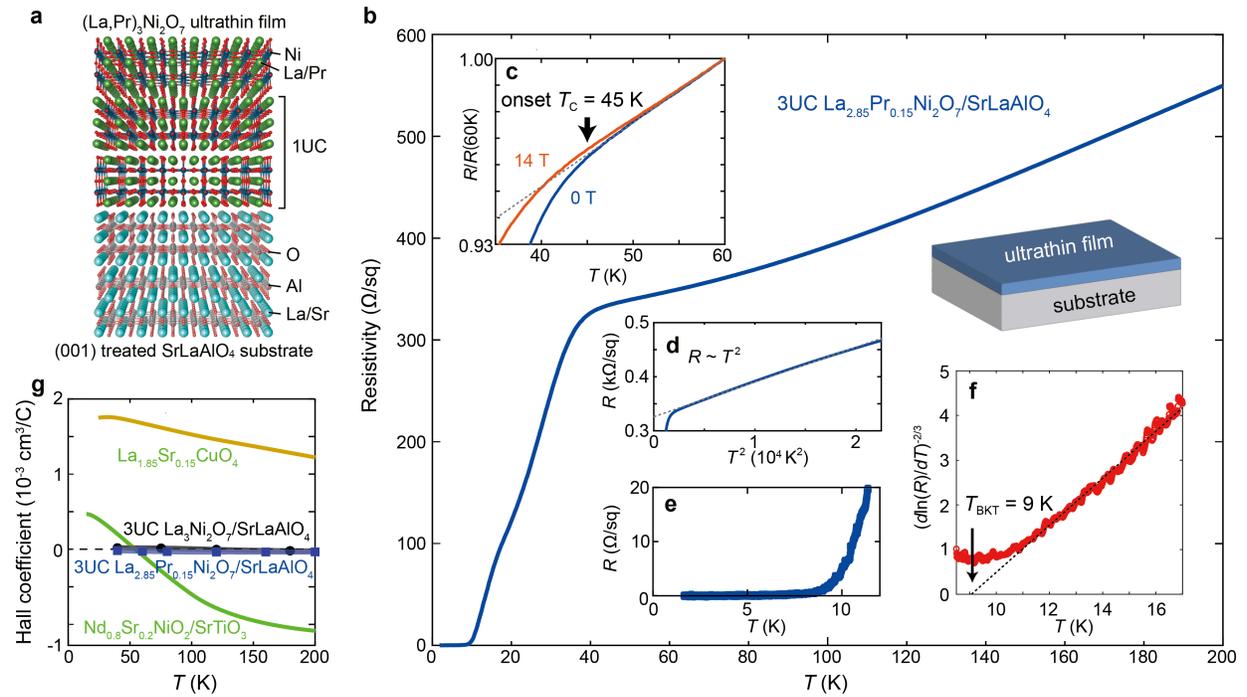

Figure 1 | Superconducting bilayer nickelate ultrathin film. **a**, The structural schematic of a (La,Pr)$_3$Ni$_2$O$_7$ bilayer nickelate film grown on treated SrLaAlO$_4$ substrate. **b**, Resistivity-temperature (*R-T*) curves for a 3UC La$_{2.85}$Pr$_{0.15}$Ni$_2$O$_7$ film on SrLaAlO$_4$. **c**, The onset $T_C$ = 45 K (down-pointing arrow) is defined as the temperature at which the *R-T* curve deviates from the linear fit in the 50–60 K range (gray dashed line) and coincides with the onset of the 14 T magnetic field response (red line). **d**, Resistivity as a function of $T^2$. **e**, *R-T* curve zoomed in near zero resistivity. **f**, $(d\ln(R)/dT)^{-2/3}$-versus-*T* curve, with linear fit in the 12–17K range extrapolated to identify $T_{BKT}$ = 9 K (down-pointing arrow). **g**, Temperature-dependent Hall coefficient for bilayer nickelate samples with and without Pr substitutions and references (La$_{1.85}$Sr$_{0.15}$CuO$_4$ and Nd$_{0.8}$Sr$_{0.2}$NiO$_2$/SrTiO$_3$) (ref.[14,41]).

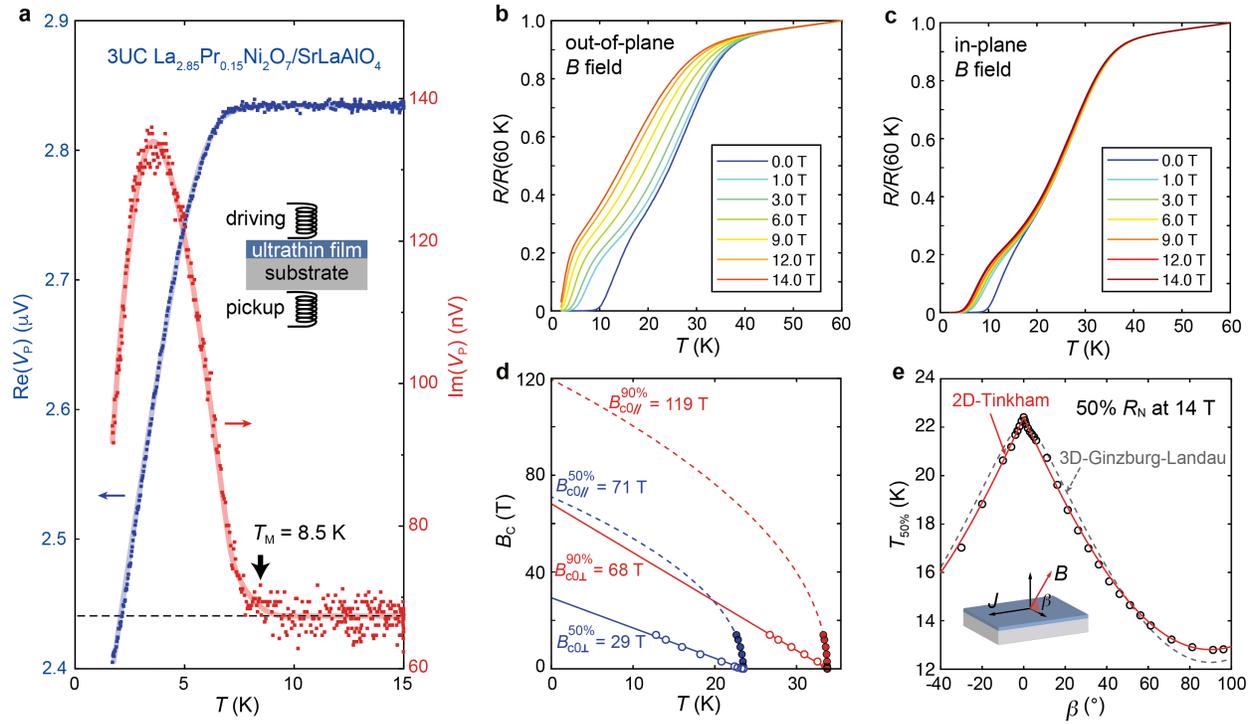

Figure 2 | Magnetic field responses of the bilayer nickelate ultrathin film. **a**, The real (Re($V_P$), blue dots) and imaginary (Im($V_P$), red dots) components of the voltage in the pickup coil measured as functions of temperature on the 3UC La$_{2.85}$Pr$_{0.15}$Ni$_2$O$_7$ film on SrLaAlO$_4$ using a two-coil mutual-inductance technique (schematic shown in inset). Shaded curves are guides to the eye. Dashed black line is a horizontal line fitting the Im($V_P$) data from 10 K to 15 K. $T_M$ = 8.5 K is determined as the temperature at which Im($V_P$) begins to deviate from the dashed black line. **b** and **c**, $R$-$T$ curves with out-of-plane and in-plane magnetic fields, respectively. **d,** In-plane (solid circles) and out-of-plane (open circles) critical fields $B_c$'s as functions of temperature. Both $B_c$'s defined by 90% (red circles) and 50% (blue circles) of normal state resistance $R_N$ are presented. Solid lines are linear fits. Dashed lines are fits using the Ginzburg-Landau formula. **e,** The temperature corresponding to 50% $R_N$ at 14 T plotted as a function of the magnetic field rotation angle $\beta$. Inset shows the measurement geometry. Open circles are measured data. Red solid and grey dashed curves are the 2D-Tinkham and 3D-Ginzburg-Landau fits, respectively.

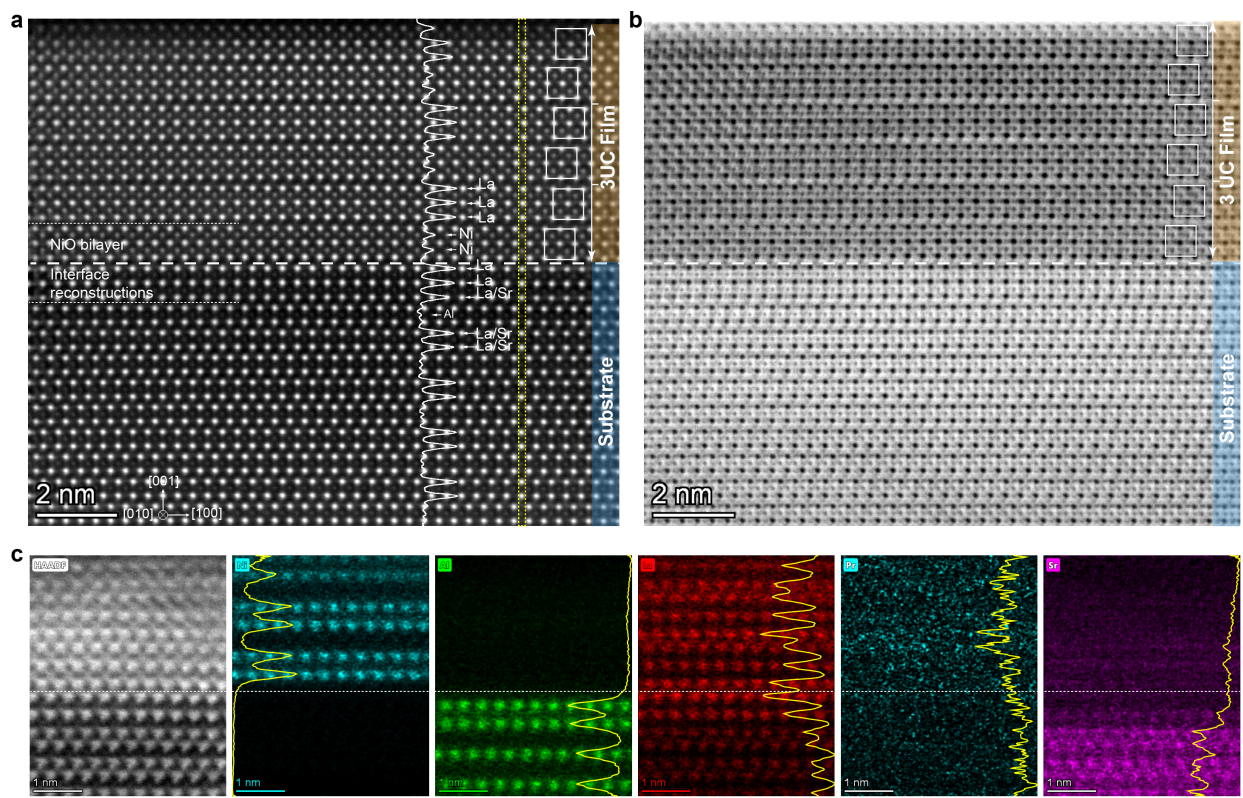

Figure 3 | Scanning transmission electron microscopy (STEM) of a superconducting bilayer nickelate ultrathin film. **a** and **b**, High-angle annular dark-field (HAADF) and annular bright field (ABF) images of a 3UC $La_{2.85}Pr_{0.15}Ni_2O_7$ film on (001)-oriented $SrLaAlO_4$ substrate. The line profile of the intensity along the yellow dashed box in **a** is shown. White squares are guides to the eye identifying the Ni-O bilayer structures. **c,** HAADF and atomically resolved energy-dispersive X-ray spectroscopy (EDS, for Ni, Al, La, Pr, and Sr, respectively) images of a same region of lattice in the same sample. The yellow curves are integrations of intensity in horizontal pixels. White dashed lines are guides to the eye indicating the position of the film-substrate interface.

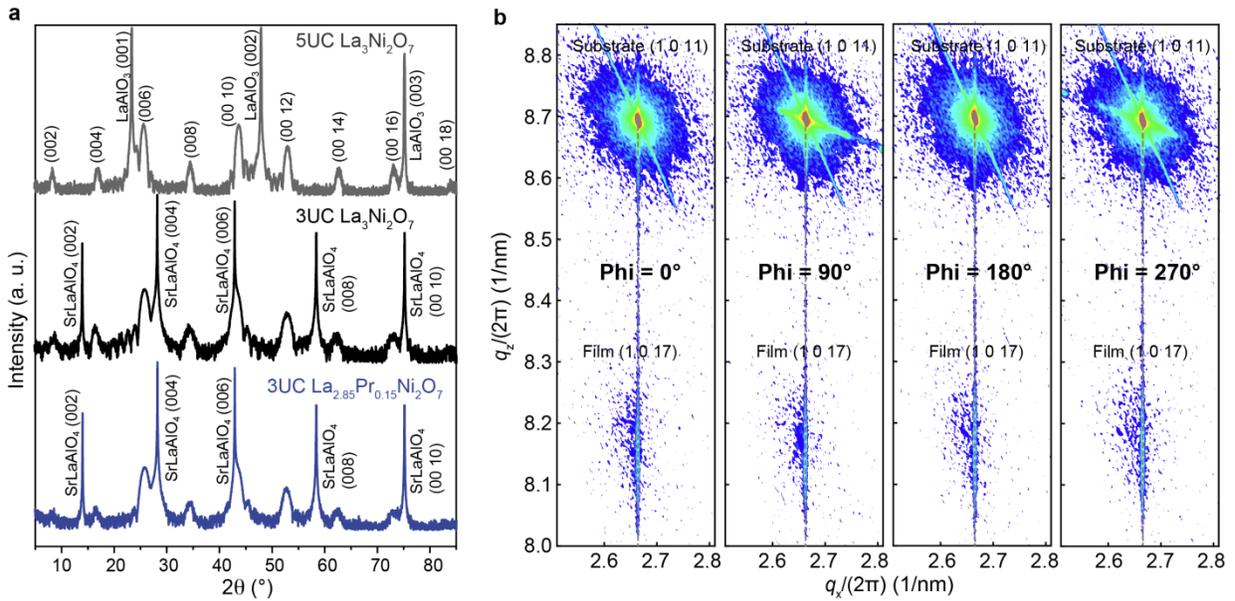

Figure 4 | X-ray diffraction (XRD) and reciprocal space mappings (RSM) of bilayer nickelate ultrathin films. **a,** XRD along the out-of-plane axis for a 5UC $La_3Ni_2O_7$ film on $LaAlO_3$ substrate, a 3UC $La_3Ni_2O_7$ film on $SrLaAlO_4$ substrate, and a 3UC $La_{2.85}Pr_{0.15}Ni_2O_7$ film on $SrLaAlO_4$ substrate. **b,** RSMs of a 3UC $La_{2.85}Pr_{0.15}Ni_2O_7$ film on $SrLaAlO_4$ along four different Phi angles. Identical $q_z$'s across the four cases reveal tetragonal phase of the film. Aligned $q_x$'s between film and substrate indicate coherent straining.

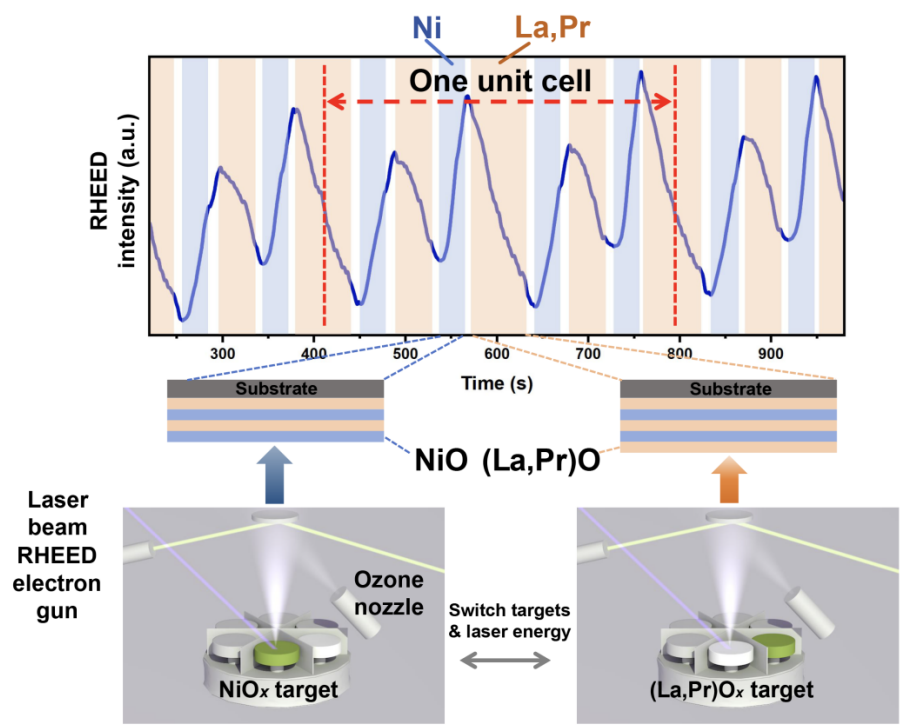

Extended Data Figure 1 | Schematic of synthesizing (La,Pr)$_3$Ni$_2$O$_7$ on SrLaAlO$_4$ with gigantic-oxidative atomically layer-by-layer epitaxy (GOALL-Epitaxy).

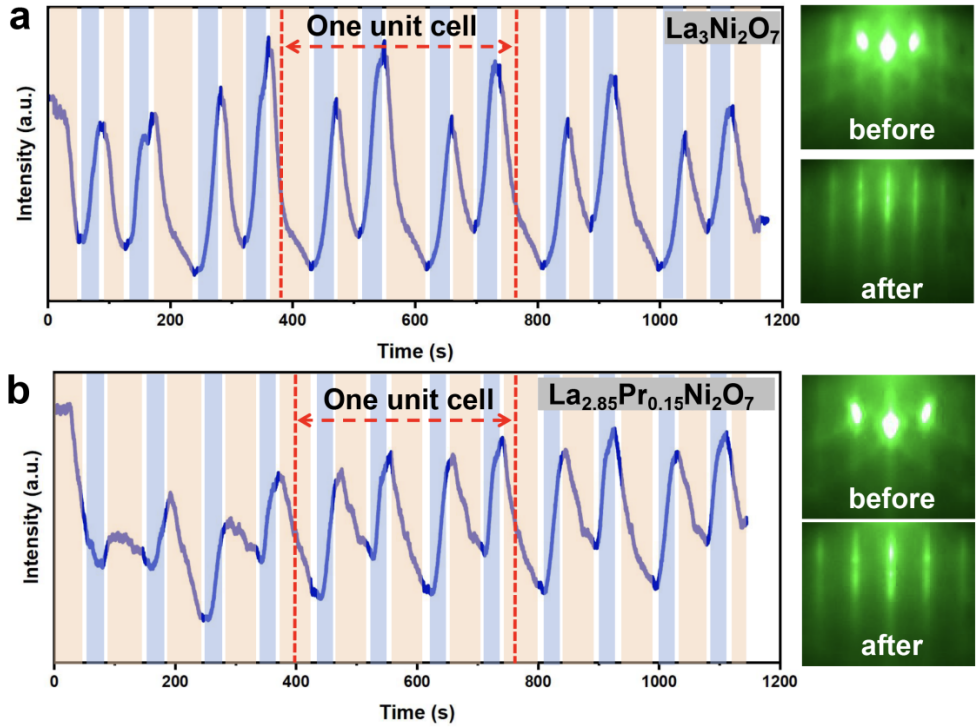

Extended Data Figure 2 | Reflection high-energy electron diffraction (RHEED) oscillations and patterns of $La_3Ni_2O_7$ and $La_{2.85}Pr_{0.15}Ni_2O_7$ growth on $SrLaAlO_4$. Blue and orange blocks represent the growth of $LaO_x$/$(La_{0.95}Pr_{0.05})O_x$ and $NiO_x$ layer, respectively.

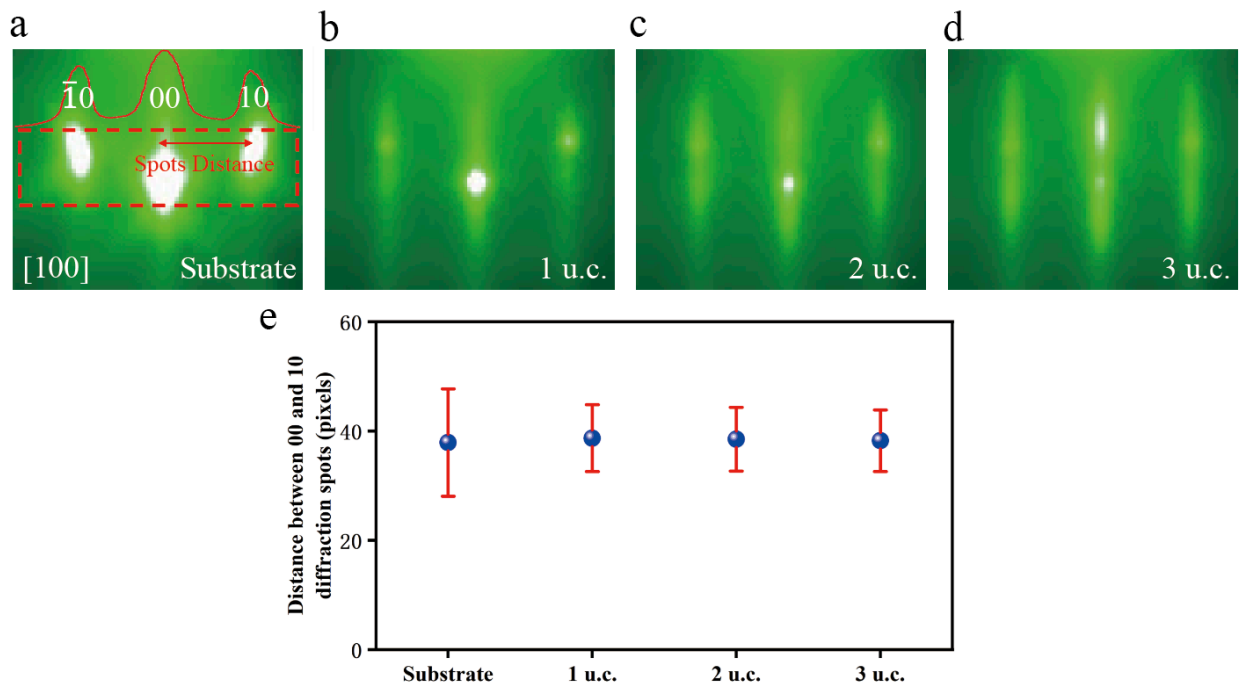

Extended Data Figure 3 | Time evolution of RHEED pattern during the growth of $La_{2.85}Pr_{0.15}Ni_2O_7$ film on $SrLaAlO_4$ substrate. **a-d**, RHEED pattern of the substrate $SrLaAlO_4$ (a), 1 unit cell of $La_{2.85}Pr_{0.15}Ni_2O_7$ (b), 2 unit cells (c) and 3 unit cells (d), respectively along [100] direction. Red solid line in (a) represents the intensity profile, which is obtained by integrating the intensity vertically in the rectangular area marked with red dashed line. **e**, Evolution of the distance between 00 and 01 diffraction spots, which is determined by the peak position of the intensity profile and the error bar is calculated by the FWHM of the peak.

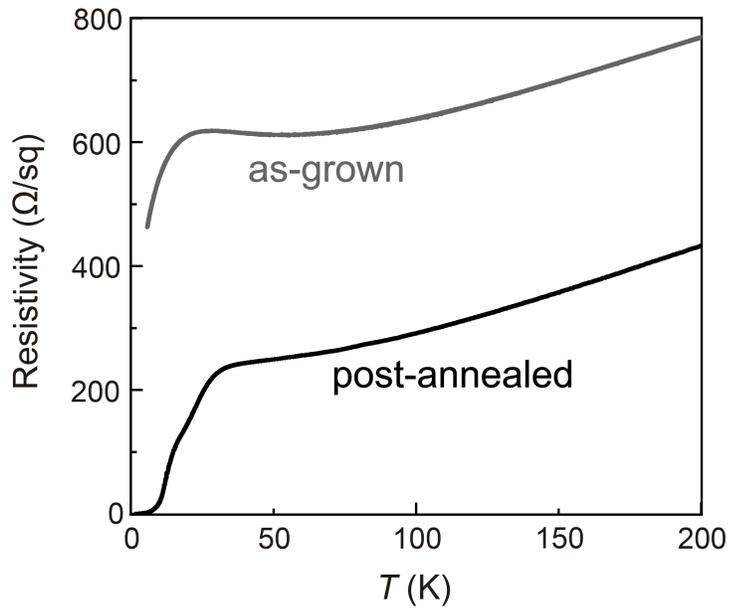

Extended Data Figure 4 | Resistivity-versus-temperature curves of a 3UC La$_{2.85}$Pr$_{0.15}$Ni$_2$O$_7$/SrLaAlO$_4$ sample before and after pure ozone flow annealing. This is a different sample from the one shown in Figure 1 in the main text.

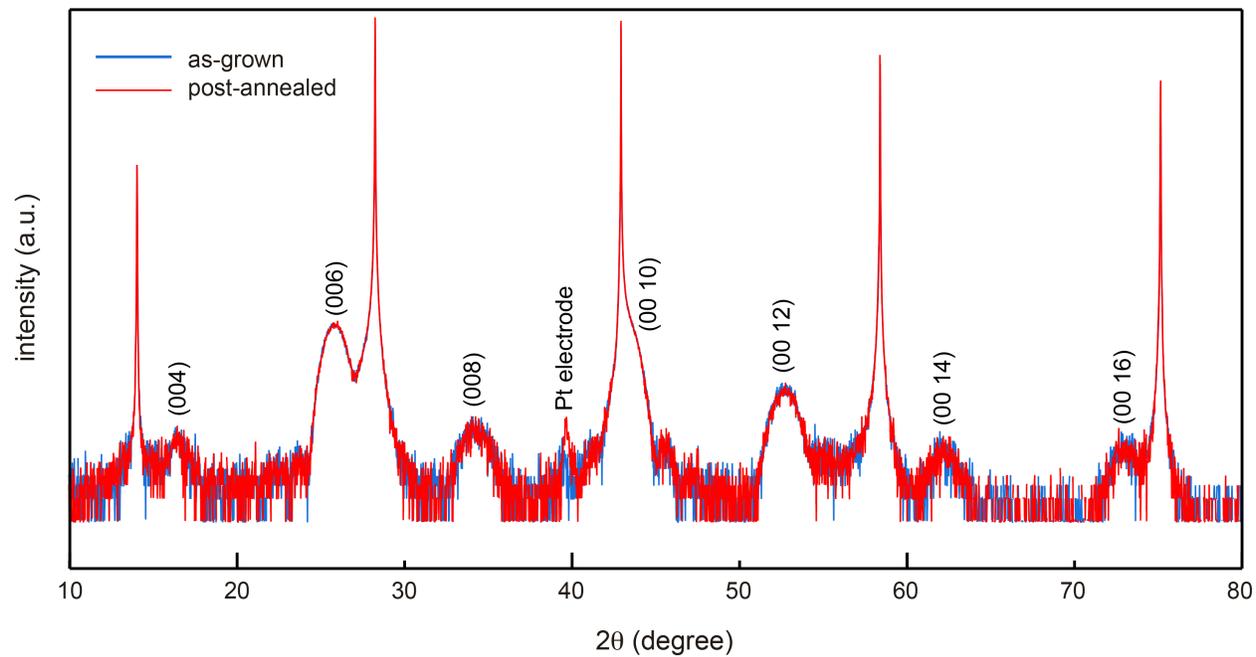

Extended Data Figure 5 | X-ray diffraction (XRD) of the same sample shown in Extended Data Figure 4 before and after pure ozone flow annealing.

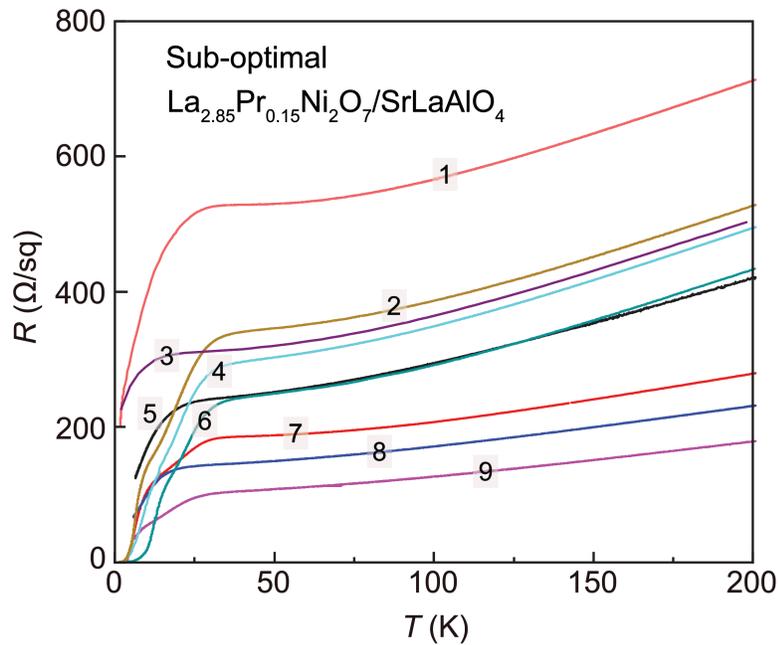

Extended Data Figure 6 | Resistivity-temperature curves for sub-optimal 3UC La$_{2.85}$Pr$_{0.15}$Ni$_2$O$_7$ films on SrLaAlO$_4$ substrates with different annealing conditions. Sample 1: Annealed at 400 °C for 30 mins, with an O$_3$ pressure of $1.57 \times 10^{-1}$ mbar. Sample 2: Annealed at 500 °C for 30 mins, with an O$_3$ pressure of $1.52 \times 10^{-1}$ mbar. Sample 3: Annealed at 700 °C for 30 mins, with an O$_3$ pressure of $1.10 \times 10^{-1}$ mbar. Sample 4: Annealed at 600 °C for 30 mins, with an O$_3$ pressure of $1.00 \times 10^{-1}$ mbar. Sample 5: Annealed at 600 °C for 30 mins, with an O$_3$ pressure of $1.38 \times 10^{-1}$ mbar. Sample 6: Annealed at 600 °C for 30 mins, with an O$_3$ pressure of $1.38 \times 10^{-1}$ mbar. Sample 7: Annealed at 600 °C for 30 mins, with an O$_3$ pressure of $1.30 \times 10^{-1}$ mbar. Sample 8: Annealed at 700 °C for 30 mins, with an O$_3$ pressure of $1.67 \times 10^{-1}$ mbar. Sample 9: Annealed at 700 °C for 30 mins, with an O$_3$ pressure of $2.03 \times 10^{-1}$ mbar. The growth conditions for all films were identical as described in the methods section.

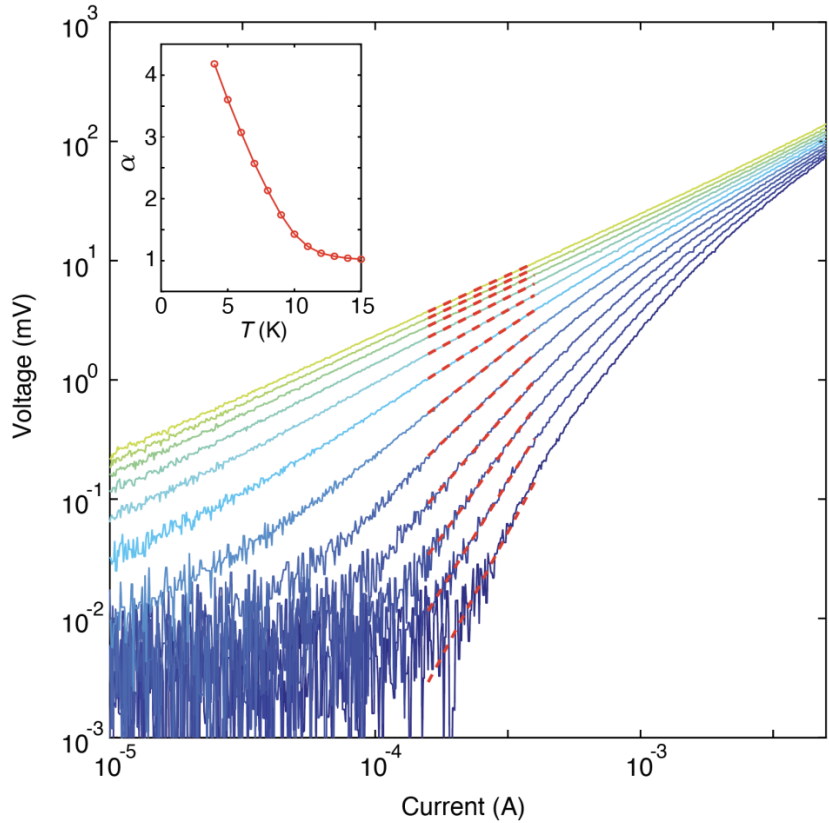

Extended Data Figure 7 | Current-voltage (*I-V*) characteristics from 4 K to 15 K, with 1 K interval. Dashed red lines are linear fits to the log-log *I-V* curves. Inset: the power law exponent $\alpha$ obtained from the fit as a function of temperature. It is important to note that, due to the significant heating effect caused by the applied current (a result of the high critical current associated with the elevated $T_C$), the actual sample temperatures are higher than the recorded values.

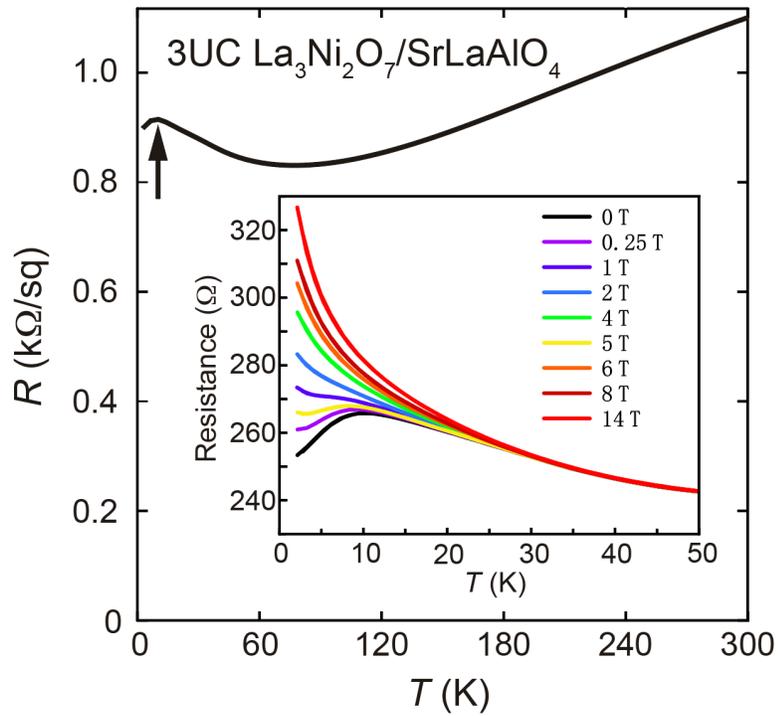

Extended Data Figure 8 | Resistivity-temperature curve and magnetic field responses (inset) for a 3UC La$_3$Ni$_2$O$_7$ film on SrLaAlO$_4$.

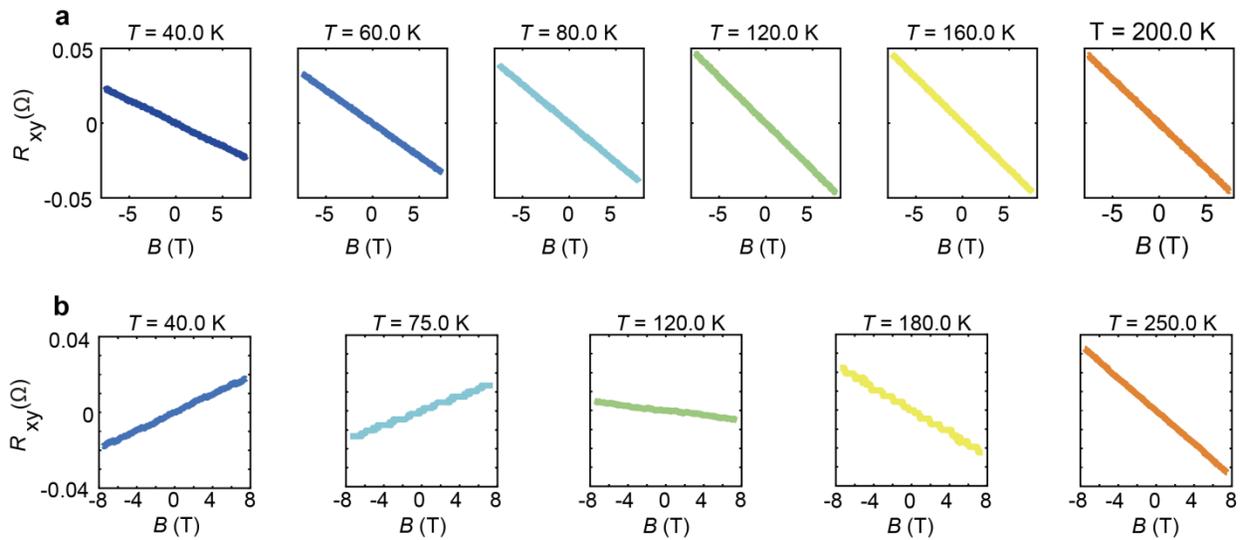

Extended Data Figure 9 | Hall effect raw data. **a,** 3UC $La_{2.85}Pr_{0.15}Ni_2O_7$ film on $SrLaAlO_4$. **b**, 3UC $La_3Ni_2O_7$ film on $SrLaAlO_4$.

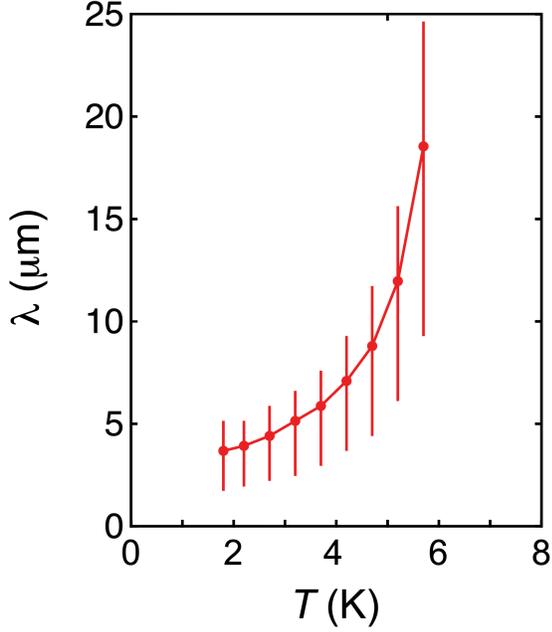

Extended Data Figure 10 | Extracted penetration depth $\lambda$ as a function of temperature from mutual inductance measurements assuming superconducting thickness of $4 \pm 3$ nm. The mutual inductance signal between two coils is given by

$$M_\infty = \pi\mu_0 \sum_{i=1}^{N_d} \sum_{j=1}^{N_p} r_{d,i} r_{p,j} \times \int_0^\infty dq \frac{e^{-qh_{i,j}} J_1(qr_{d,i}) J_1(qr_{p,j})}{\cosh Qd + \left(\frac{Q^2 + q^2}{2qQ}\right) \sinh Qd}$$

where $N_d$ and $N_p$ represent the total turns of driving coil and pick-up coil respectively. $r_{d,i}$ and $r_{p,j}$ represent the radius of the $i$-th and $j$-th coil, $h_{i,j}$ represents the distance between the $i$-th and $j$-th coil. $d$ is the thickness of superconducting film $Q^2 = q^2 + \lambda^{-2} - i\mu_0\omega\sigma_1$, where $\lambda$ is penetration depth and $\sigma_1$ is conductivity of quasiparticles. The flux-leak is lower than 2% in our measurement setup as calibrated by a 5mm×5mm×0.5mm Nb plate under 2K.

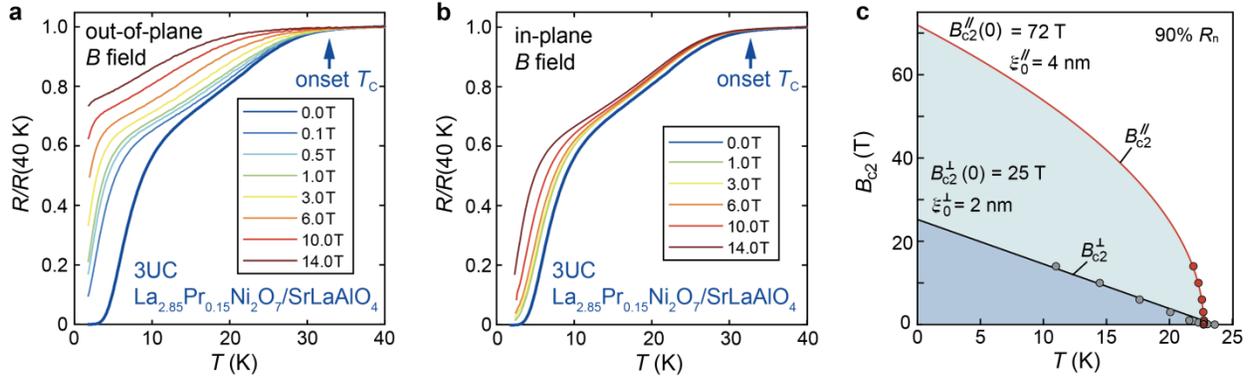

Extended Data Figure 11 | Two-dimensional superconductivity of an onset-$T_C$-32-K sample. **a** and **b**, Out-of-plane and in-plane magnetic field responses for a 3UC La$_{2.85}$Pr$_{0.15}$Ni$_2$O$_7$ film on SrLaAlO$_4$, respectively. **c,** In-plane and out-of-plane upper critical field $B_{c2}$ as a function of temperature. $B_{c2}$ is defined by the magnetic field needed to reach 90% of normal state resistance $R_n$. Coherence lengths are calculated with the equation $\xi_0 = \sqrt{\phi_0/2\pi B_{c2}}$.

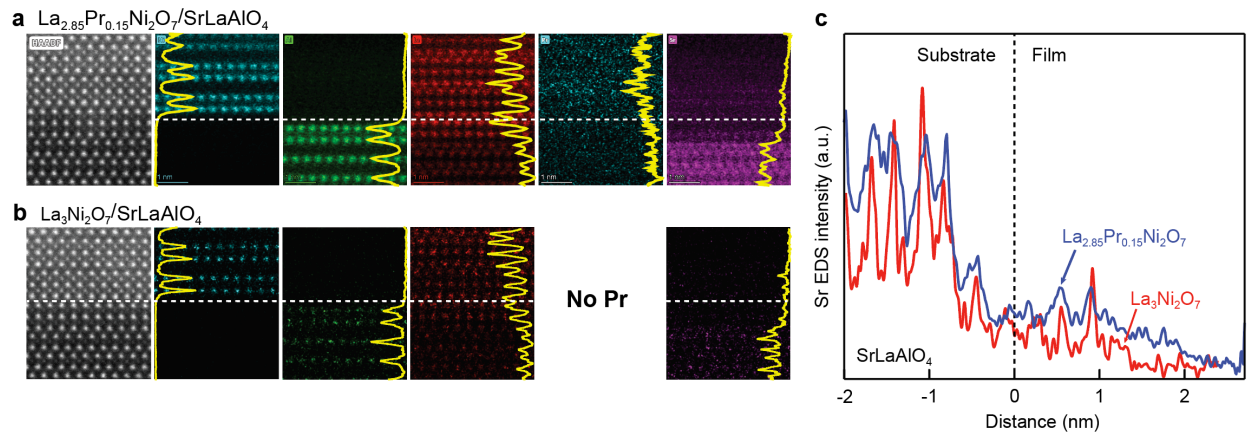

Extended Data Figure 12 | STEM HAADF and atomically-resolved EDS of an 3UC $La_{2.85}Pr_{0.15}Ni_2O_7$/$SrLaAlO_4$ superconducting sample (**a**) and a 3UC $La_3Ni_2O_7$/$SrLaAlO_4$ sample (**b**). **c**, Sr EDS intensity for both samples as a function of distance across the interface.

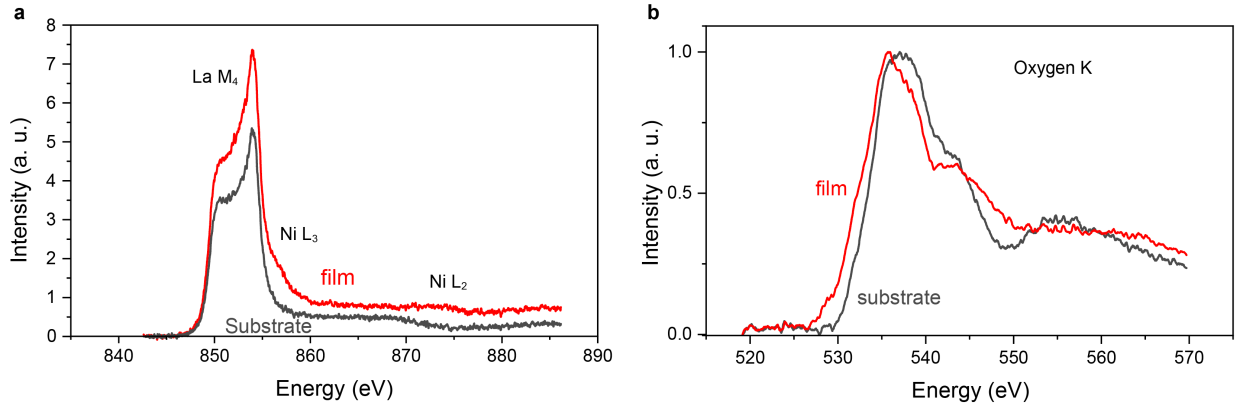

Extended Data Figure 13 | Electron energy loss spectroscopy (EELS) of Ni L (a) and O K (b) edges, comparing film and substrate, in an annealed 3UC La$_{2.85}$Pr$_{0.15}$Ni$_2$O$_7$/SrLaAlO$_4$ superconducting sample.

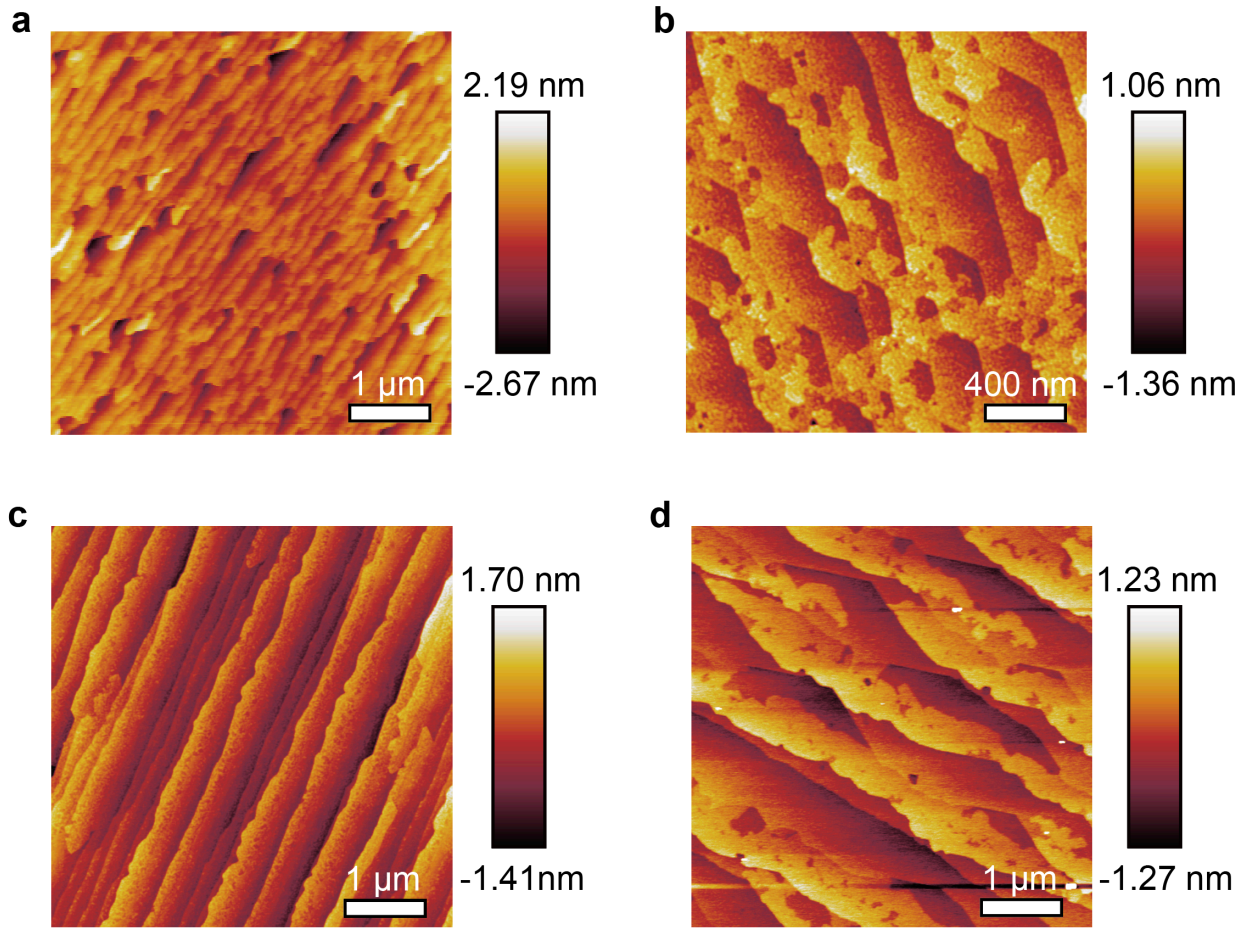

Extended Data Figure 14 | Atomic force microscope (AFM) images. **a**, **c** AFM image of the SrLaAlO$_4$ substrate annealed at 1030°C and 1080 °C, respectively. The root-mean-square roughness is 536.6 pm and 439.6 pm, respectively. **b**, AFM image of La$_3$Ni$_2$O$_7$ film on SrLaAlO$_4$ substrate. The root-mean-square roughness is 308.6 pm. **d**, AFM image of La$_{2.85}$Pr$_{0.15}$Ni$_2$O$_7$ film on SrLaAlO$_4$ substrate. The root-mean-square roughness is 407.3 pm.

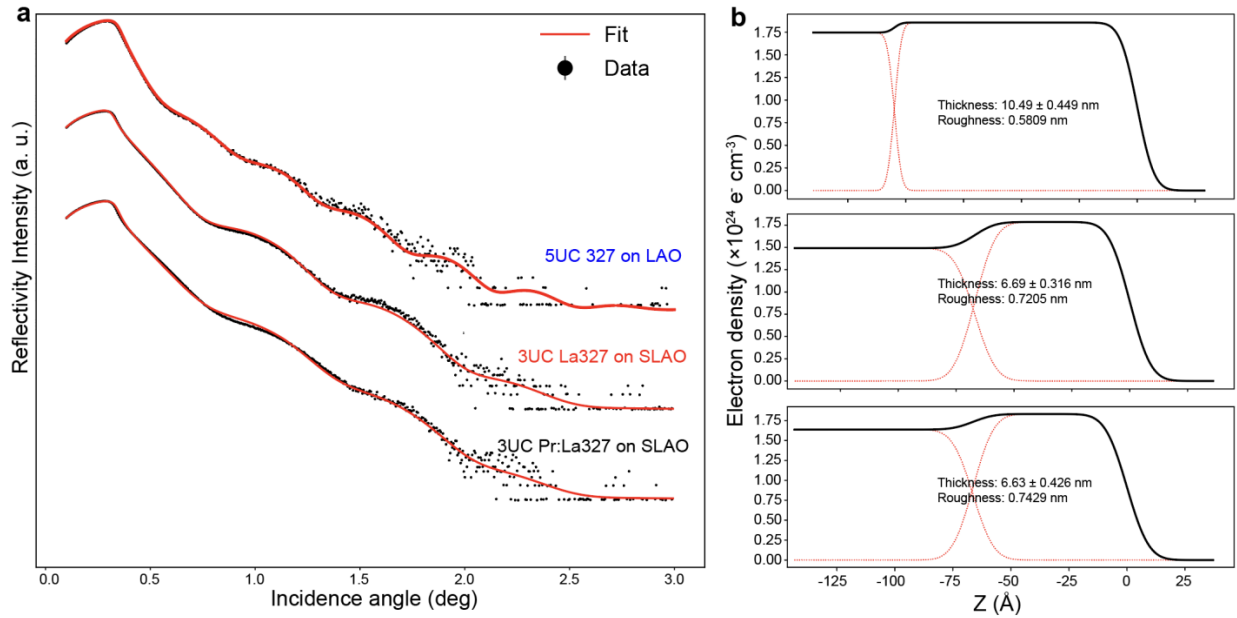

Extended Data Figure 15 | Thickness and roughness analysis by X-ray reflectivity (XRR). **a**, XRR of 5 unit cell $La_3Ni_2O_7/LaAlO_3$ (top), 3-unit-cell $La_3Ni_2O_7/SrLaAlO_4$ (middle) and 3-unit-cell $La_{2.85}Pr_{0.15}Ni_2O_7/SrLaAlO_4$ (bottom) films. **b**, The relationship between electron density and thickness obtained by fitting the XRR results. The film thickness and surface roughness obtained by fitting are also listed.

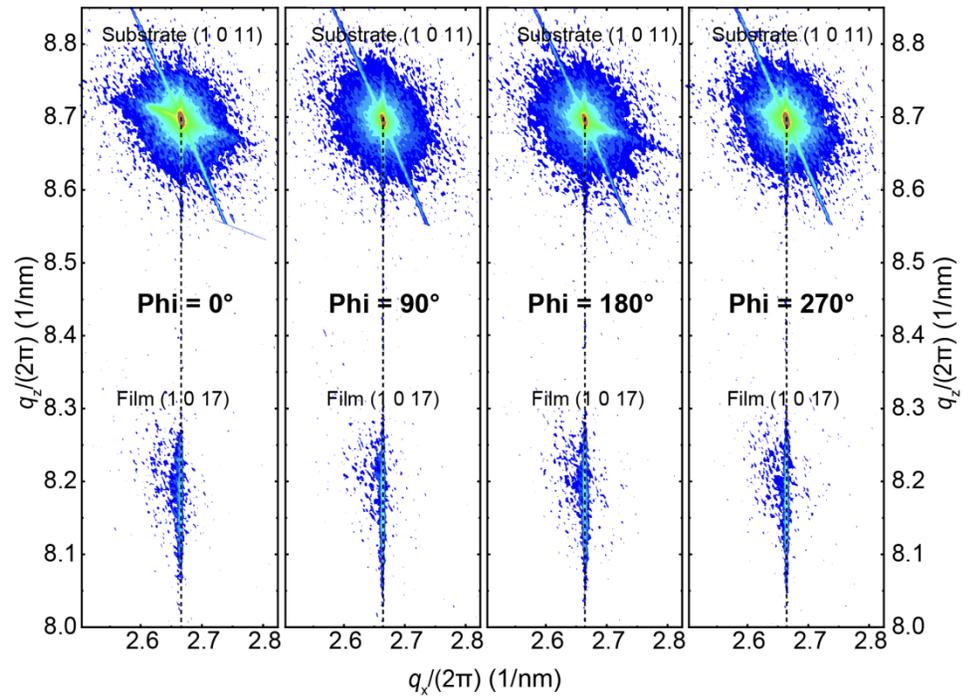

Extended Data Figure 16 | RSMs of a 3UC $La_3Ni_2O_7$ film on $SrLaAlO_4$ sample along four different Phi angles.

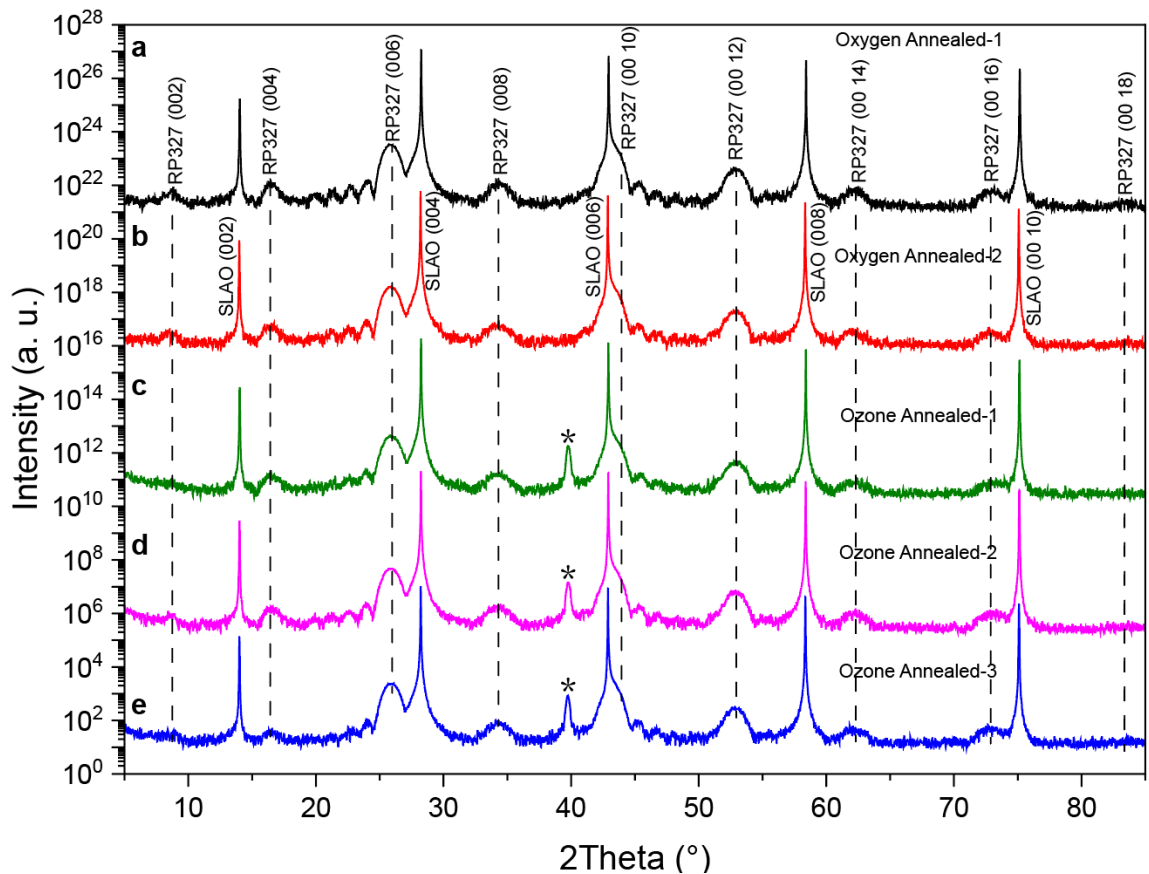

Extended Data Figure 17 | X-ray diffraction θ-2θ symmetric scans of 3-unit-cell $La_3Ni_2O_7/SrLaAlO_4$ after annealing. **a-e** is the XRD pattern of the same sample after annealing in oxygen atmosphere and ozone, respectively. The dashed lines in the figure demonstrate that the diffraction peak has remained essentially unchanged after several cycles of annealing, indicating the preservation of a stable crystal structure. This implies that XRD is no longer adequate for detecting the minor oxygen deficiency that is pivotal in inducing the superconducting transition. The asterisk comes from the contribution of the electrode. "RP327" stands for Ruddlesden-Popper 327 phase.

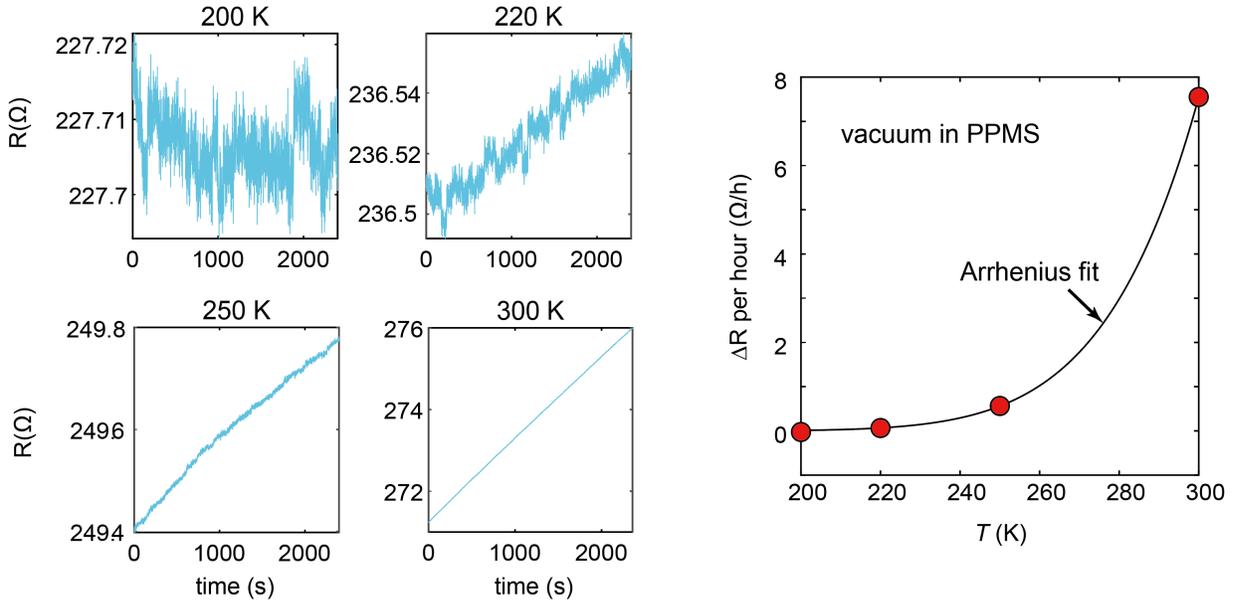

Extended Data Figure 18 | Resistance changes as functions of time, with sample maintained at vacuum and different temperature inside the quantum design physical property measurement system (PPMS). An Arrhenius fit yields an activation energy of 0.34 eV for the oxygen loss process.

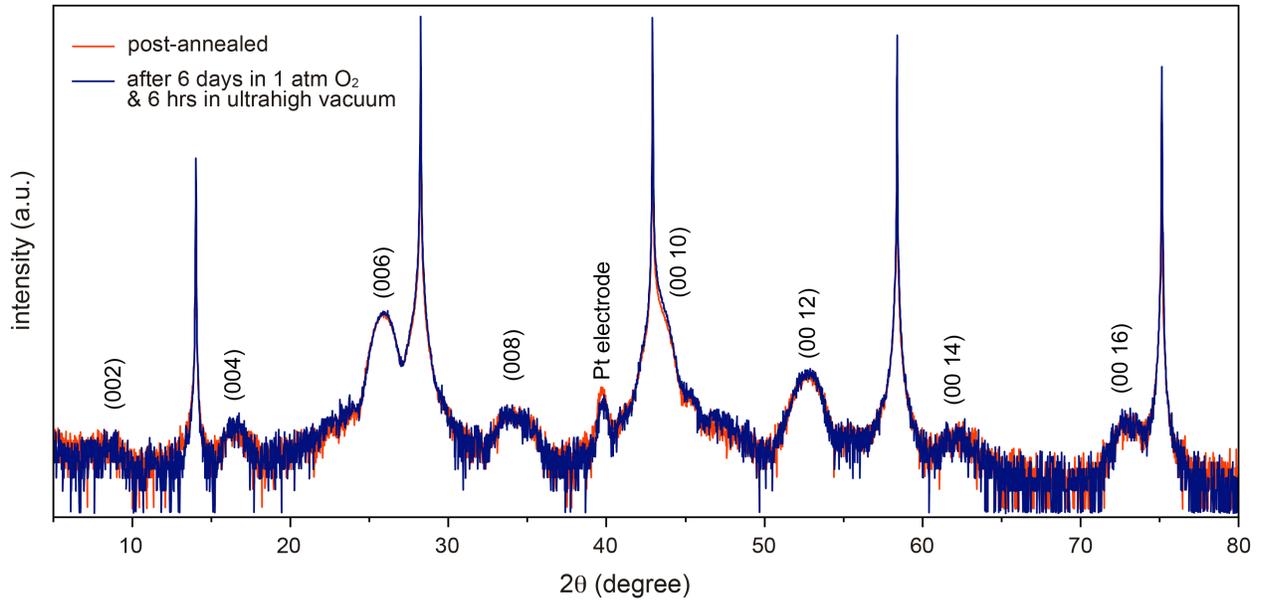

Extended Data Figure 19 | XRD of the same sample shown in Figure 1 in the main text after post-annealing after 6 hours in ultrahigh vacuum and 6 days in 1 atm $O_2$ and at ambient temperature.